# Enabling a Multi-Purpose High-Energy Neutron Source Based on High-Current Compact Cyclotrons

*Lance L Snead[1,2], Daniel Winklehner[1], Dennis Whyte[1], Steve Zinkle[3], Zach Hartwig[1], David Sprouster[1,2]*

1) Massachusetts Institute of Technology (Cambridge, MA)
2) Stony Brook University (Stony Brook, NY)
3) University of Tennessee-Knoxville (Knoxville, TN)

## 1. OPPORTUNITY AND CONCEPT SUMMARY

The current and future need for high-energy neutrons has been a subject of increasing discussion and concern. Immediate applications for such an intense neutron source include medical isotope production, high-energy physics (HEP) research, and for materials development and to support qualification for fission reactors. Also, and of the utmost importance, is the need for such a source to inform critical gaps in our understanding of the transmutation materials science issues facing fusion power reactors.

For the case of medical isotope production there has been simultaneous pressures advocating new neutron sources including the increased need for therapeutics, the loss and pending loss of the current fission reactors used for their manufacture, and the need to engage in domestic production of important isotopes such as Mo-99. The fission reactor community has similar concerns regarding the loss of the aging materials-test-reactor fleet, comprised of facilities averaging more than a half-century in age. Moreover, the current national and international lack of *fast* reactors has hampered our ability to understand irradiation-induced damage and to provide qualification-level data on materials for the array of fast reactors currently under development.

A 14 MeV fusion prototypical neutron source (FPNS) has been a critical, yet unresolved need of the fusion program for more than 40 years. Given the narrowing timeline for construction of pilot and fusion power plants the urgency and necessity of such a neutron source has become increasingly time sensitive. This is a consensus opinion as highlighted in the 2002 Greenwald report [3] through the recent NASEM [4] and FESAC [5] reports. As originally envisioned, the deuteron-on-lithium target (D-Li) technology suggested by Grand [6] was developed by the FMIT-2 program and facility [7]. Originally slated for 100 mA, FMIT encountered numerous issues achieving current and accelerator duty cycle [8] an yielded only very low dose data. In the years since abandoning FMIT-2 the community has undertaken the IFMIF/EVEDA program: an activity developing technologies consistent with a facility merging two RFQ/LINACs onto a flowing Li target. That system would provide 250 mA of 40 MeV D-on-Li. The IFMIF/EVEDA projected duty cycle aims for 65-70%.[9] While many accomplishments (technology requirements) have occurred as part of the IFMIF/EVEDA project, the anticipated IFMIF system (see Fig 1) cost has increased to ~2 Billion USD making it essentially fiscally prohibitive.

Recently, the Office of Fusion Energy challenged the US fusion community to envision a scaled-down yet scientifically and technically relevant and economic FPNS, thus potentially providing the US with a world leading capability to reinforce its current leadership in fusion material and technology. The result of that study was to suggest three competing technologies, each with essential trade-offs in technology readiness levels (TRL) and economics:

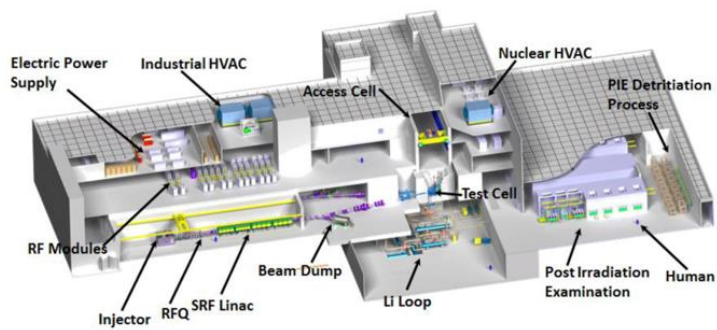

*Figure 1. Artists concept of the IFMIF facility. [1]*

- *IFMIF-Lite*: A scaled-down version of IFMIF technology operating at 125 mA. Beam and target technology to mimic (leverage) that developed under the IFMIF/EVEDA program.
- *Spallation Neutron Source*: Tailored to closely simulate the fusion neutron spectrum. A 800 MeV, 1MW $p^+$ beam delivered by the LANSCE accelerator with a predicted 18 dpa/yr [10].
- *D-T Gas Target Design*: Based on Phoenix Nuclear Laboratory technology of D-D and D-T neutron generators. 14 MeV neutrons through 80 mA D beam incident on high pressure T gas.

As reviewed, all three concepts may provide neutrons relevant for fusion. However, major questions whether they can provide neutrons reliably, economically, and at relevant annual fluence exist. As presented, the *IFMIF-Lite* and *Spallation Neutron Source* are both at relatively high-TRL, though arguably have high capital and operating costs and the deviation from fusion-prototypic transmutation and pulsed effects of the spallation source are troubling to the materials science community. In contrast, the *D-T Gas Target Design* is intriguing but efforts to increase neutron flux fall more than an order of magnitude short of that necessary. [11] For these reasons a more economic neutron source which is of similarly prototypic transmutation characteristics to IFMIF is sought.

Within this white paper, a blueprint of necessary R&D to enable a transformational change in both the capital and operating cost of the IFMIF-Lite **driver** concept is presented. Enabling this transformation is the replacement of the historic RFQ/LINAC components with <u>multiple compact 35+ MeV D drivers.</u> Through use of multiple drivers, we not only bring this application into the range of current high-availability commercial cyclotrons, but also gain the technical and operational advantages of utilizing multiple independent beams on target. Commercial templates for these normal conducting cyclotron drivers currently exist, requiring modest and near-turn development in *subsystems* to achieve our goal of a 50-100+ mA upgradeable driver configuration. Figure 2 provides images of the IBA Cyclone 70P and the Best 70P Copper compact cyclotrons which have been installed at a number of facilities and provide high availability 70MeV (equivalent to 35 MeV D) protons at ~1 mA. While the proposed subsystem development work is based on these nominally 70MeV compact Copper machines, the technology developed is largely independent of choice of Cu-based or Superconducting-based (SC-based) cyclotron. While the proposed high-current enhancements will be immediately applicable to these Cu-machines, they will also translate to the higher-risk, potentially higher-reward superconducting machines which are under development.

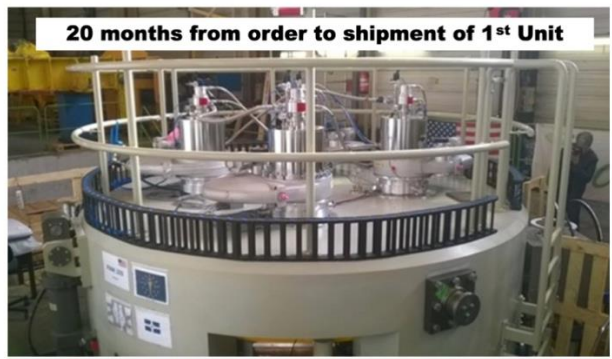
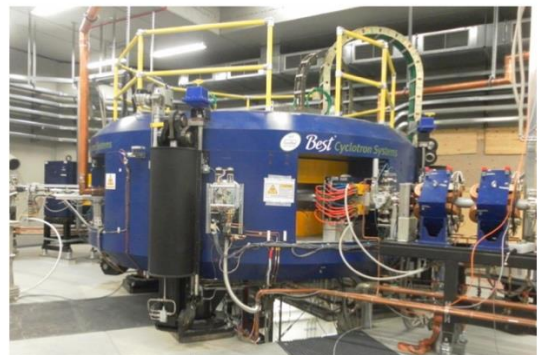

*Figure 2: Currently installed compact Cu-based medical isotope cyclotrons of 70 MeV (H)/35 MeV (D)*

The ultimate performance of this compact-fusion prototypic neutron source (C-FPNS) improves upon the Linac/RFQ concept through a combination of technical improvements over the current medical isotope cyclotron technology coupled with an operational scenario whereby a fixed number of units are continually beam-on-target enabled by a scheduled rotation of instruments in and out of a maintenance rotation. In this way the C-FPNS facility can avoid single-fault shutdown caused by the driver(s), allowing availability to be maximized and dictated by the non-driver systems. The number of total units would be determined by the achievable current and machine availability proven as part of this proposed work, noting that the industry-standard medical isotope cyclotron availability currently exceeds 92% as compared with the assumed 65-70% availability of the IFMIF system. Moreover, this assumed operational scenario allows maximum flexibility for facility uprate by rotating-in higher fluence or potentially higher energy cyclotrons as they come available.

*Within the period of this proposed research a number of critical beam-related technological risks will be retired to enable the use commercial cyclotrons as high-flux neutron sources. The technical achievements allowing such a breakthrough will enable several practical and scientific applications including medical isotope production, fission materials damage research, high-energy density physics research, and the long-sought solution to a 14 MeV fusion simulator: realizing a near-term compact FPNS (C-FPNS). R&D will be carried out in close collaboration with commercial vendors who will provide build-ready drawing for the first-of-a-kind C-FPNS driver. The driver will then be commercially built and undergo testing to determine parameters of beam current and availability over long periods while being beneficially used to determine/confirm important design parameters associated with the Deuterium->Lithium target.*

## 2. STATUS OF TECHNOLOGY AND NEEDED NEAR-TERM ADVANCEMENT

This proposed compact cyclotron driver solution has performance *and* economic advantages compared to RFQ/LINAC driven alternatives driven by the following attributes: i) lower projected cost-for-current, ii) compactness (<2m radius superconducting, <3 m normal conducting); iii) elimination of external cryogenics; iv) reduced installed power requirements, and v) a more flexible and efficient operating scenario. In order to understand how these advantages may be harnessed and demonstrated near-term an understanding of the current state of commercial cyclotron technology is needed. This is outlined in the following sections. It is noted that this is an ideal time to execute on this technology development plan given the existence of a program and team actively engaged in very similar research to achieve a high dose compact cyclotron at our desired energy and current, albeit with a hydrogen beam as compared our desired deuterium beam.

2.1 Current Commercial Scale Cyclotrons

The use and variety of cyclotrons for medical isotope production is widespread with over 150 units operating in the US and over 1200 world-wide. The range of therapeutic isotopes is quite broad and the physical space constraints of commercial cyclotrons range from relatively unconstrained for large isotope production facilities to very constrained in the case of hospital settings. Table 1 provides a snapshot in the types of cyclotrons currently on the market for various applications.

| Application | Typical User | Maximum Energy (MeV) | Maximum Current (mA) |
|---|---|---|---|
| **Proton Therapy** | Hospital | 200-250 | $10^{-6}$ |
| **Radioisotope Production/Research** | Research Lab | 70 | 0.5-0.7 |
| **SPECT Radioisotope Production** | Research Lab/Industry | 30 | 0.5-1.3 |
| **PET Radioisotope Production** | Hospital/Industry | 15-25 | 0.1-0.4 |
| **PET Radioisotope Production** | Hospital | 10-12 | 0.05 |

*Table 1: Comparing the application and specifics of commercial cyclotrons.*

The majority of modern medical isotope production cyclotrons (Cu-based) accelerate negatively charged ions such as H$^-$ and/or D$^-$, with leading commercial suppliers including GE Healthcare (US), IBA(Belgium),

BEST Cyclotron Systems (Canada), Advanced Cyclotron Systems (Canada), and Siemens CTA(Germany). Of these, the larger companies that manufacture both Copper and superconducting machines include GE Healthcare and IBA. The leading manufacturers dedicated solely to superconducting cyclotrons include Mevion Medical Systems and Ionetix. A direct comparison of Cu-based, and SC-based machines is not straightforward, though a basic comparison showing size and ion energy is found in Figure 3. In this Figure the more compact nature of the superconducting machines, enabled by the higher magnetic field achievable with the superconducting magnets, is evident. In this comparison the research-grade machines of Antaya were selected, though there are many examples of commercial workhorse machines such as the Ionetix 12C 12MeV compact superconducting cyclotron utilized for N-13 production. This Ionetix machine has a relatively modest current (~0.025 mA) with both an internal ion source and target. Machines such as this typically operate in hospital laboratories at night to produce short-lived isotopes that are harvested and used during the day. A different class of superconducting cyclotron is the Mevion Medical Systems, which has a large installed base of 250 MeV proton accelerators for proton therapy. As with all proton therapy accelerators the required current is low, in this case the Mevion 250 is rated at 100 nA.

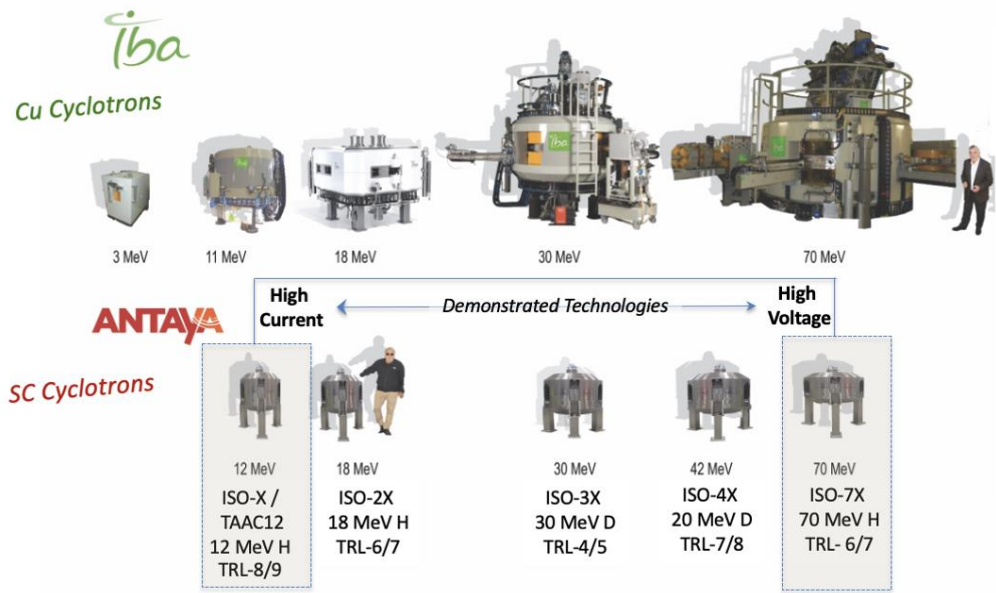

*Figure 3: Scale comparison of Cu and Superconducting cyclotrons.*

The closest analog cyclotron to that needed for a compact cyclotron driver application are the specialized isotope (SPECT) producing cyclotrons, which are typically in the 30 MeV range and produce up to 1 mA proton current, though are also rated for use with deuterium. These machines are high-availability workhorses that produce, as example, imaging isotopes such as [111]In and [99m]Tc. The industry standard availability (beam-on-target) for these machines, of which examples are shown in Figure 2, is >92%. They are typically (as case of Figure 2) axially *injected* external ion source machines (meaning the beam is injected from outside of the cyclotron downward from the top-center of the machine as compared to *horizontal injection* which introduces the beam through the median plane.) These are also *external* target machines (meaning the beam exits the cyclotron vessel and is carried to the target) as compared to the *internal* target machines typified by the compact hospital-use cyclotrons. As mentioned above, the larger Cu-based machines of Figure 2 can serve as a blue-print for a C-FPNS machine as the cyclotron beam physics of the 70 MeV proton machines very closely maps to the 35 MeV deuterium beam desired for a C-FPNS. However, and as will be touched on in the remainder of this section, with solutions provided in the following sections, the current must be significantly enhanced from the nominal 1 mA current level to >5 mA, and that enhanced current must be managed into a stable beam configuration within the cyclotron and extracted from the machine with an acceptable level of loss.

As will be discussed, such high-current cyclotrons have been designed and are being demonstrated [12, 13] that are consistent with the nominal 35 MeV deuterium machines of Figure 2, though will require R&D to operate with deuterium. An ongoing program to deliver a similar machine to our desired C-FPNS driver current exists. These efforts come out of the US High Energy Density Physics IsoDAR cold neutrino program, which is developing a 10 mA, 60 MeV proton driver as an academia led team closely collaborating with IBA. Their ultimate goal is to enable a definitive search for so-called *sterile neutrinos* within 5 years of data-taking. [14] The main challenge and risk being retired by IsoDAR is directly in-line with our needs, the development of a high-current injection source, the mitigation of resulting space charge effects (the electric self-field pushing the particles away from each other) and an efficient cyclotron-physics management extraction of that beam from the cyclotron.

To address space-charge challenges, the IsoDAR collaboration has developed the HCHC-XX design, on which we can build for this $D^+$ program. HCHC stands for High Current $H_2^+$ Cyclotron, and XX for beam energies from 2 MeV/u to 60 MeV/u. The HCHC-XX is a modular cyclotron design that – in addition to high energy gain per turn (due to its four double-gap RF cavities driven at 32.8 MHz) and a $\nu_r = 1$ precession resonance to increase turn separation at the end – uses three novel concepts that make the project high-risk/high-return:

- Acceleration of $H_2^+$ instead of protons (5 mA of $H_2^+$ will yield 10 mA of protons later on)
- Direct axial injection through a radio-frequency quadrupole (RFQ) [13, 15, 16]
- Utilization of so-called *vortex motion* (a collective effect of high-current beams) [12]

The following subsections will discuss these three aspects and how we can leverage the $H_2^+$ program for high-current $D^+$ beams. In short, a high-current beam ($H_2^+$ or $D^+$) is injected into the cyclotron, carefully controlling beam losses and achieving maximum inter-turn separation between the final two orbits with low beam halo, which would otherwise contaminate the space between these orbits. An electrostatic extraction system is inserted, further increasing the inter-turn separation until magnetic insertion devices can be used to guide the beam out of the machine. The IsoDAR team has adopted a safety limit of 200 W (a relative loss of ~1 in 10,000 particles at 5 mA and 60 MeV/u) of continuous beam losses during extraction, which is based on the operational experience with the Injector II cyclotron at the Paul Scherrer Institute in Switzerland [2, 17]. A schematic view of an HCHC-01 and an HCHC-60 is shown in Figure 4.

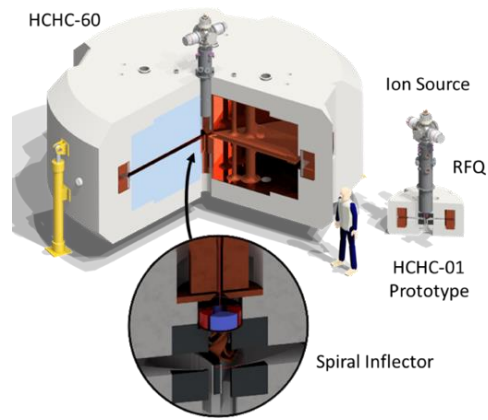

*Figure 4: Examples of an HCHC-60 and an HCHC-01. The $H_2^+$ is produced in the ion source with 80% purity, pre-accelerated and bunched in the RFQ, and guided into the cyclotron through a spiral inflector with electrostatic transverse focusing.*

2.2 High Current Injection ($H_2^+$ and $D^+$)

The ability to move from ~1mA to >5mA cyclotron beams is the central technical hurdle to achieving the C-FPNS with the key developments concerning the injection system. Requirements for that system are that it centers the beam with respect to the central region of the cyclotron, matches the beam phase-space within the cyclotron eigenellipse, further centers the beam with respect to the median plane, and provides both longitudinal and radially matching within acceptable beam loss. Such high-current axial-injectors have been designed and are being demonstrated [12, 18] that are consistent with the nominal 35 MeV deuterium machines of Figure 2, though will require R&D to operate with deuterium. Specifically, the HCHC-XX design features acceleration of molecular hydrogen ions ($H_2^+$) because the binding electron can be stripped by guiding the beam through a thin carbon foil, yielding 2 protons per $H_2^+$ ion. Thus, 5 mA of $H_2^+$ can be accelerated that yield 10mA of protons, alleviating some of the space-charge related challenges during injection. The needed primary $H_2^+$ beam is produced in the MIST-1 ion source, designed, built, and operating at MIT [19, 20]

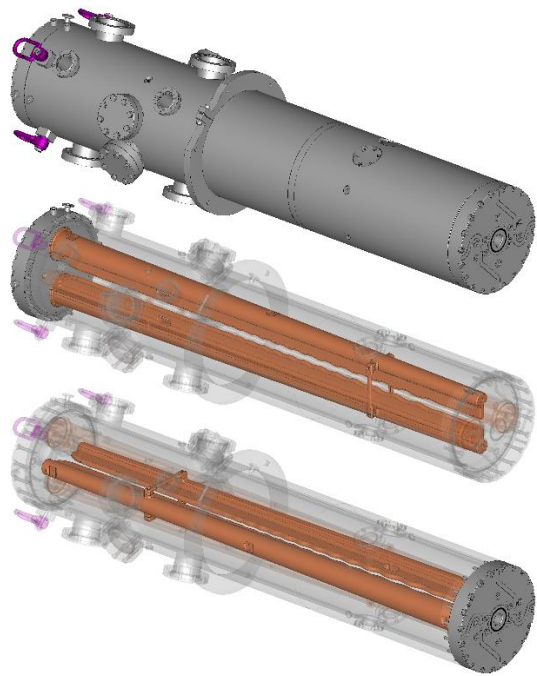

Figure 5: The HCHC-XX Radio-Frequency Quadrupole (RFQ).

In any particle accelerator driven by radio-frequency (RF) electromagnetic waves only a narrow *phase window* can be populated per RF period. Particles that arrive before or after this window will not be properly accelerated, which leads to energy spread, reduced beam quality, and beam loss. Hence, to further reduce space-charge in the injector beam line, the HCHC-XX design uses an RFQ to efficiently bunch the beam before guiding it on the cyclotron's median plane through a so-called *spiral inflector*. This RFQ (see Figure 5) is of the *split-coaxial* type, allowing it to have a small diameter despite its low resonant frequency of 32.8 MHz (corresponding to the cyclotron RF frequency). It is only 1.25 m long and accelerates to a modest 70 keV. Because the highly bunched ions quickly begin to spread (partly due to their high space charge) in both the transverse and longitudinal directions, the RFQ exit is brought as close to the median plane as possible to begin the acceleration process in the cyclotron and utilize the strong focusing of the Azimuthally Varying Field (AVF). Nevertheless, transverse refocusing using an electrostatic quadrupole device is needed before the ions can enter the spiral inflector. The use of an RFQ embedded in the cyclotron yoke for direct axial injection is a novel technique that has not been used before. The IsoDAR collaboration performed a number of high-fidelity simulations to test the feasibility of this design, including Uncertainty Quantification using Statistical and Machine Learning [19]. To experimentally test this new injection scheme, a prototype RFQ is currently under construction and tests injecting the beam into a dipole magnet and appropriate spiral inflector are funded and planned for 2023.

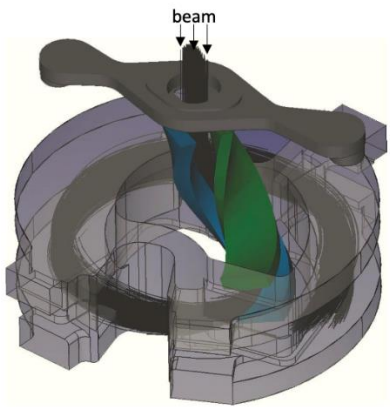

*Figure 6: The spiral inflector, bringing the axially injected beam onto the median plane (horizontal) of the cyclotron.*

The RFQ-Direct Injection Project (RFQ-DIP) will increase injection efficiency from ion source to turn 6 by a factor of five, conservatively

Page **6** of **17**

estimating the injection efficiency to ~50%, based on simulation. We can compare this, for example, to PSI Injector II [17], where 25 mA of proton beam current are required from the ion source, to achieve 2.2 mA extracted from the cyclotron. The main losses happen before and during injection, to prepare and clean up the beam. This constitutes < 10% injection efficiency. Similarly, the industry leader IBA offers the Cyclone 70 XP, with a 10 mA H$^-$ current from the external ion source and a guaranteed extracted proton current of 750 µA. Again, a < 10% injection efficiency [21] [22]. Here, stripping extraction is employed.

The ability to move from ~1mA to >5mA cyclotron beam is the central technical hurdle to achieving the C-FPNS with the key developments concerning the injection system and a secondary challenge being clean extraction (see following subsections). As described above, the HCHC-XX design has made it possible, at least on paper, to achieve these goals for H$_2^+$/protons. However, experimental demonstration is a strict requirement for such a high-risk/high-reward concept. Consequently, an injection test bench is currently under construction for demonstration end of calendar 2023.

Because of the same charge and almost the same mass of H$_2^+$ and D$^+$, all concepts of the HCHC-XX are directly applicable for a D$^+$ high-current cyclotron and lessons-learned can be applied. From a standpoint of retiring risks, it seems most prudent to follow this suggested path to demonstrate the feasibility of a compact high-current D$^+$ cyclotron (HCDC):

1. Design and fabricate a 2MeV/u test cyclotron to expand the H$_2^+$ study from injection to injection + acceleration for 6 turns. Here H$_2^+$ has the distinct advantage of a much higher Coulomb threshold for nuclear effects, making radioprotection easier.
2. For D$^+$, the RFQ should be slightly modified. Size and RF considerations will not change. The vane modulation will be optimized for the small difference in mass of the two ions.
3. Procure a D$^+$ ion source (readily available off-the-shelf) and duplicate the IsoDAR RFQ. Because the technical design is complete, fabrication of a second machine will be cost-effective.
4. Combine the D$^+$ source with the new D$^+$ RFQ and the test cyclotron from *1* above. For 6 turns, no major modifications are required, and the RF frequency adjustment is within the tuning range.

In this scheme, H$_2^+$ ion source and RFQ1 will be used in the IsoDAR cyclotron and D$^+$ ion source and RFQ2 will be available for a full 35 MeV/u, 5 mA D$^+$ prototype as shown in Figure 7.

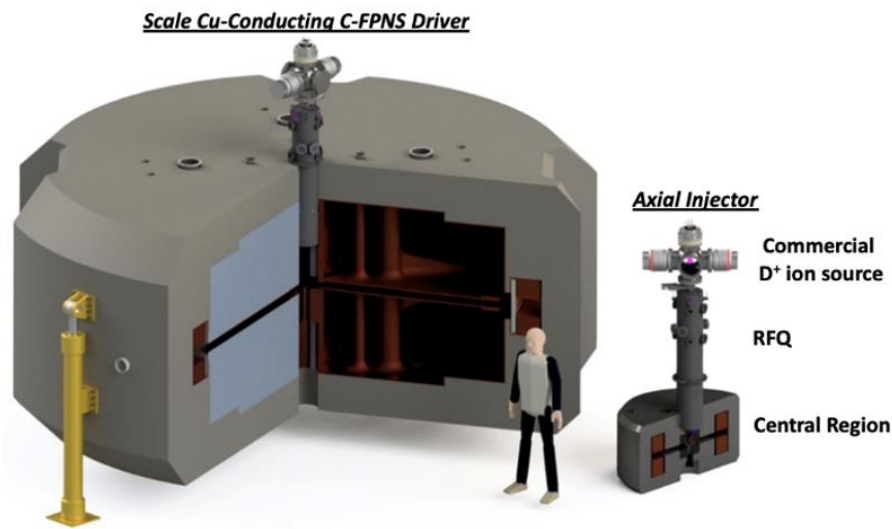

*Figure 7: Right: D+ injection and acceleration test. Left: Resistive C-FPNS Driver.*

## 2.3 Beam Collimation

Once inside the cyclotron, the bunches experience the external force of the cyclotron magnetic field and the internal force from space charge. Normally, space-charge would be a negative effect, but in an isochronous cyclotron, the combination of these two forces can lead to longitudinal-radial coupling, and – if the beam current, bunch size, and external fields are matched – a circular motion of the bunch around its own axis in the bunch rest-frame, forming a stable vortex with length and width being approximately equal. This effect was first observed in PSI Injector II [2] and subsequently reproduced in high-fidelity Particle-In-Cell (PIC) simulations using OPAL [23]. Theoretical explanations can be found elsewhere [19, 24, 25] with the brief summary being that in addition to the spiral orbit through the cyclotron during acceleration, the individual ions also experience an $\vec{E} \times \vec{B}$ drift (see Figure 8), leading to the "vortexing." If the injected bunches are short enough, this effect yields a stable, round, steady-state distribution that has a net longitudinal and radial focusing effect. PSI Injector II is the only cyclotron in operation where this effect is used, and the HCHC-XX is the first cyclotron actively designed to use vortex motion for clean extraction.

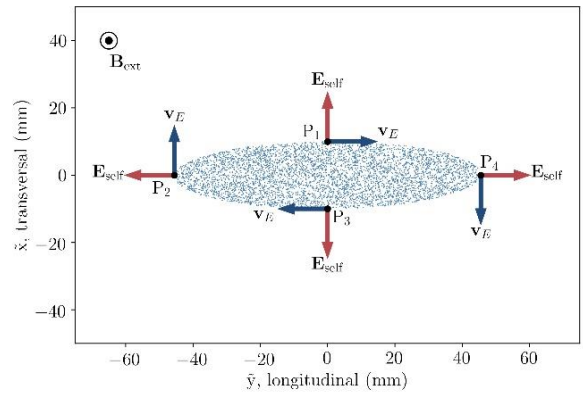

*Figure 8: A bunch in its local frame (following the beam orbit in the cyclotron). Additional velocity components arising from the combination of electric self-fields due to space charge and the cyclotron magnetic field are indicated.*

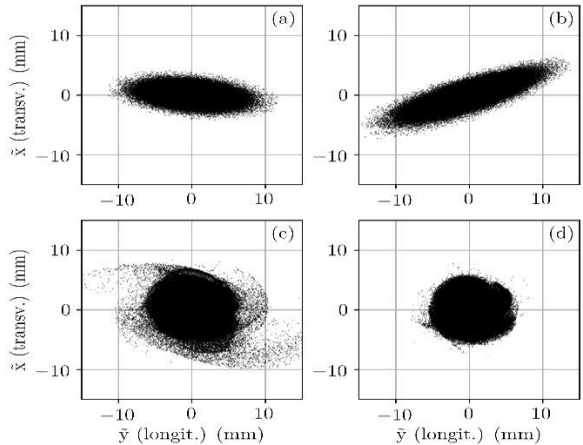

*Figure 9: (a) initial beam after the spiral inflector, (b) beam after 7 turns without space charge (for comparison), (c) beam after 7 turns with space charge (6.65 mA), (d) beam after 7 turns with collimators in place.*

For the HCHC-XX design, the IsoDAR collaboration has optimized the cyclotron and beam injection using the OPAL code and placed collimators in the first four turns of the cyclotron to remove particles (halo) that forms while the vortex is established. Due to the spiral inflector, the initial beam injected into the central region is not well matched to the ideal beam within the cyclotron. The beam is large in the vertical direction ($\tilde{z}$) and has large emittance. Again, using the development of the IsoDAR accelerator and injector as an example [19] local frame projection of the beam onto the median plane ($\tilde{x} - \tilde{y}$) is shown in figure 9 (a). As a comparison, a beam without space-charge is shown in figure 9 (b). This beam is long and will "propeller", causing radial overlap of adjacent turns at certain azimuths. With the proper space charge in the simulation, after seven turns, the stationary (matched) distribution has formed as can be seen in figure 9 (c). The halo that has formed in the process is cut away by placing 12 collimators (see figure 10). This beam loss does not lead to activation, as the particle energy is still below 1.5 MeV/amu. Combined with the RFQ transmission, the total beam loss from ion source to 60 MeV/amu is only about 40% in the HCHC-XX design, which is more than a factor 5 lower than other reported beam losses of high-current injection machines. By using this technique, the turn separation at 60 MeV/u is maximized and the number of halo particles in between turns is minimized. Using the high-fidelity OPAL simulations, it was shown that for the HCHC-XX, the beam loss in the extraction region is around 100 W, a factor two below the safety limit (see next section). As mentioned earlier, the safety limit of 200 W is based on operational experience at PSI, where they see up to 10 mSv/h in the extraction beam line of Injector II.

For a $D^+$ cyclotron, the collimation scheme would need to be adapted to collimate earlier, reflecting the lower Coulomb threshold. Common materials in the cyclotron are OFHC copper, 304 Stainless Steel, 1010 Mild Steel, Vanadium Permendur, 6061-T6 Aluminum, and Tungsten. It will be necessary to carefully evaluate the losses both in the central region during collimation (typically water-cooled copper), during acceleration (here the choice of RT vs. SC becomes especially important, as beam losses can heat the cold-mass of the SC cyclotron) due to gas-stripping and Lorentz-stripping, and finally, of course in the extraction channel, where the beam has reached its highest energy.

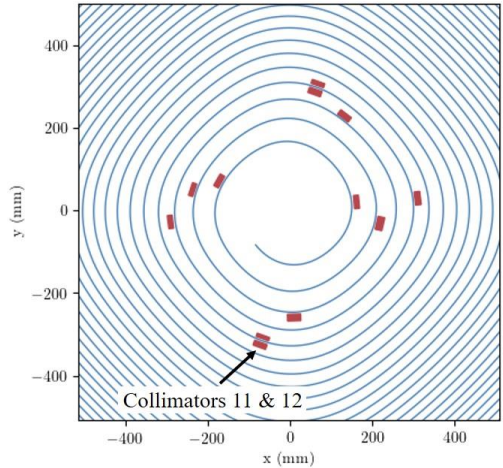

*Figure 10:12 collimators placement in the first turns of IsoDAR HCHC-60 cyclotron.*

2.4 Extraction

As discussed, a C-FPNS driver will by necessity be an external target machine common to the copper machines such as those of Figure 2. As the cyclotron magnetic configuration is essentially designed as an ion trap, the extraction (moving the beam outside of the last internal orbit and placing it outside of a magnetic field and delivering it along a path to an external target) is challenging. Typical methods for extraction include (1) the use of stripping foils which have a wide array of practical advantages including simplicity and essentially 100% extraction efficiency, and (2) a combination of electrostatic and magnetostatic extraction channels. Both options exist, for example, for the IBA Cyclone-70 XP: ($H^-$/$D^-$) strippers and ($^4He^{2+}$/$H_2^+$) electrostatic deflection. Stripping is feasible in this case, because of the still fairly low currents extracted from the Cyclone-70 XP (750 µA $H^-$). However, C-FPNS driver currents would lead to high heat deposition, and foil failure would likely be a huge challenge. In addition, beam

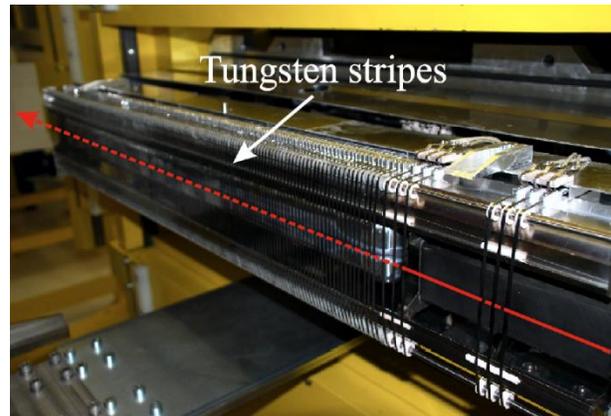

*Figure 11: The electrostatic extraction system at PSI Injector II. The spring-loaded tungsten stripes are ~0.05 mm thick and form the septum that is inserted between turn N-1 and turn N. From [2]*

losses due to Lorentz stripping in a SC cyclotron would also be challenging, as would the increased activation of the machine due to gas stripping.

In the HCHC-60, electrostatic extraction is used, similar to PSI Injector II, which is the highest performing proton cyclotron in the world. In the HCHC-60, at 60 MeV/amu, the losses during the extraction process may not exceed 200 W when hands-on maintenance is occasionally required. Otherwise, the radioactive activation of the machine would be too high. This limit was empirically determined at PSI for the operation of their Injector II cyclotron. To achieve this relative particle loss of $10^{-4}$, very clean beam extraction is needed and can be achieved by maximizing inter-turn separation of beam paths and minimizing beam halo (particles that potentially reside between turns). To maximize inter-turn separation, the four double-gap RF cavities can provide an energy gain per turn of almost 2 MeV in the final turn. In addition, a precessional resonance from the fall-off of the magnetic field increases turn separation locally at the position of the electrostatic extraction system comprising a thin, grounded septum inserted between turn N-1 and N, and a negative electrode pulling the beam outwards. The PSI Injector II extraction system is shown in figure 11 and the placement of these electrostatic septa is schematically shown in figure 12. Their function is to create a radially outwards pointing electric field of 8-10 MV/m deflecting the beam over a long path during its final turn. This technique provides enough space to subsequently install a magnetic septum guiding the beam out of the machine. The IsoDAR collaboration showed the required low beam losses at extraction through high fidelity simulations using particle-in-cell (PIC) code OPAL [26]. The optimized design was tested for robustness using a statistical learning method called *surrogate modelling based on polynomial chaos expansion*. Typical variations of RF phase and -frequency, magnetic field, etc. were also considered. In [13]the robustness of the design was shown.

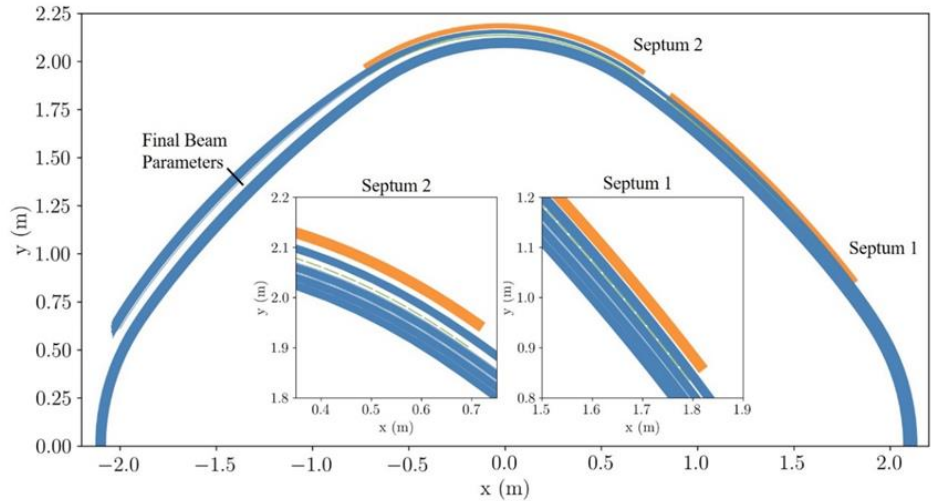

*Figure 12: Top view of the last four turns in the HCHC-60 simulated with OPAL. The placement of two electrostatic extraction channels (Septum 1 and Septum 2) is shown, as well as the effect of the 3D electric field generated by them on the beam.*

A similar extraction scheme to the HCHC-60 is likely the most feasible option for a high current $D^+$ cyclotron. Here, simulations should be repeated for $D^+$ to understand where the beam loss hot-spots in the cyclotron and during extraction are, and what materials are involved (101 copper for the RF cavity, tungsten for the septum, etc.). Geant4 or similar codes should then be used to understand activation in the respective places.

2.5 Parallel Operational and Design Issues Defining C-FPNS Driver Design

The potential utility of compact cyclotrons as C-FPNS drivers reside both in the technical performance of the cyclotrons and the operational advantages allowed through use of multiple drivers. Beyond demonstrating that the cyclotron-based C-FPNS system can meet or exceed the performance metrics set forth by the community[27] the economic attractiveness of the system becomes paramount. These potential economic advantages are interconnected at the system/facility level such that the cyclotron development must be carried out in parallel with certain operational considerations.

Cyclotron Footprint

As the building and facilities supporting the classis RFQ/Linac are on the order of 40% of the total facility cost, and the required square footage is quite large [28] (4 acres green field accommodating a 35,000 sqft accelerator facility building, 10,000 sqft shop and operator building, and 6,000 Li building) the physical space requirements for a ring of accelerators surrounding he lithium target is a central cost and potentially a large savings. The *footprint* cost will be affected by the ultimate choice of normal or superconducting systems, whether there is a need for separate shop space for individual maintenance, and importantly what is the level of activation of the machine (potentially from collimation and extraction losses) that would potentially require external operational and/or maintenance shielding. As discussed in sections 2.2 and 2.3 the combination of beam physics from collimation and beam extraction with materials choices will define both maintenance and routine operational shielding requirements.

Neutron Economy of Multi-Beam Target

The classic RFQ/Linac-based IFMIF assumed two driver beams of 40 MeV-125 mA converging on a nominally 25 mm flowing lithium wall. In the reduced-performance DONES HFTM (high flux test module) only one beam of the same performance was assumed[29], providing ~0.3 liters of volume at approximately 10 dpa/year. While this is a similar displacement rate as defined in the US FPNS metrics[27], the volume is a factor of 6 larger. Some idea of how DONES or the reduced energy IFMIF-Proto (ENS) can be gained from Figure 13, which provides a studies summary of target station volume-versus-dpa (in iron). However, it is noted that the US metrics defines a considerably lower-volume system and a simple linear scaling from the DONES is not appropriate, as incident angle, deuterium energy, incident beam width, and acceptable flux gradient are determining factors. As seen in the lower left-hand portion of Figure 13, the maximum damage (dpa) occurs near the face of the target assembly closest to the lithium jet, with a significant flux gradient radially from the mid-point of that surface. This gradient becomes stronger and the central flux magnified as the impinging beam is narrowed, as in this example by moving from 20x5 $cm^2$ to 10x5 $cm^2$ beam impacting the lithium wall.

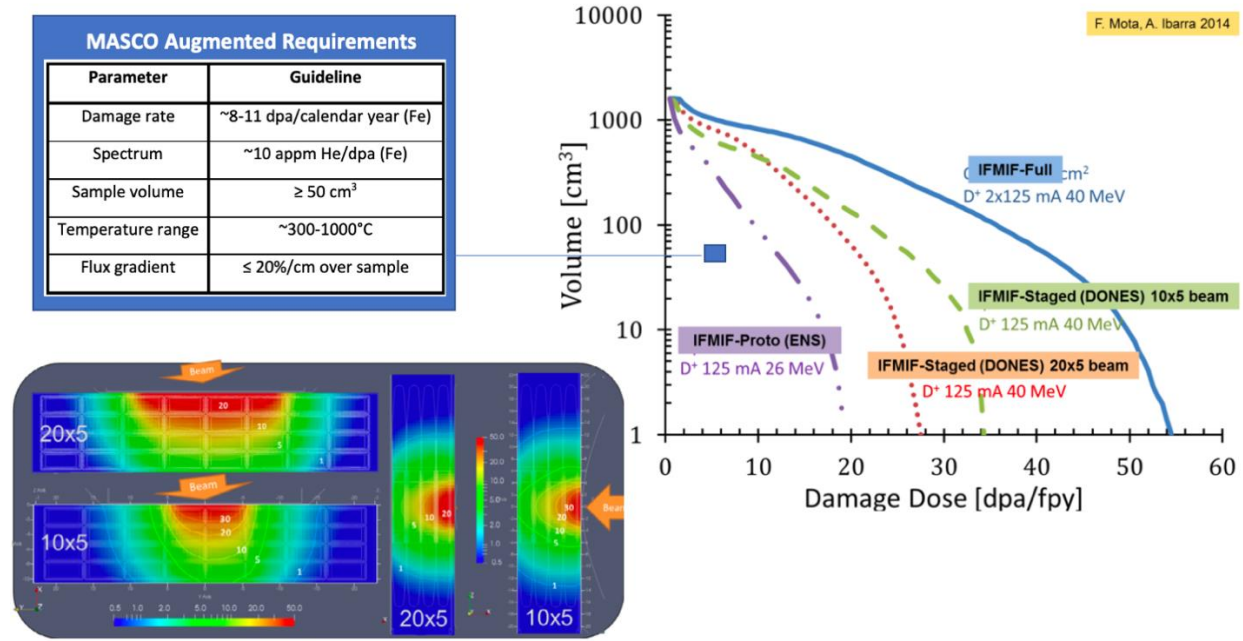

*Figure 13: The effect of high flux test module (HFTM) volume on damage dose for various studied systems (after Mota) as compared to the US-FPNS Augmented Requirements [27]. On lower left DPA isocontours of the target assembly for a DONES-like beam (125 mA, 40 MeV.) [29]*

The opportunity presented through the use of multiple target beams is of *beam-flattening* through changing the incidence angle of the beams away from the single nearly-normal beam. This will have the effect of reducing the gradient and making more effective (maximizing) use of the flux over the high flux test module. This is shown conceptually in Figure 14, which provides an exaggerated case of three beams of the same total current as the IFMIF-Proto Beam of Figure 13. In this example, over a nominally 5cm x 10cm x depth box the effect of utilizing three beams is to reduce the flux gradient over the box from 75% (the case of the IFMIF-Proto) to 25%. As this gradient will in part determine the placement of the HFTM a related increase in flux (or decrease in required current and cost) can be realized and assumed in design.

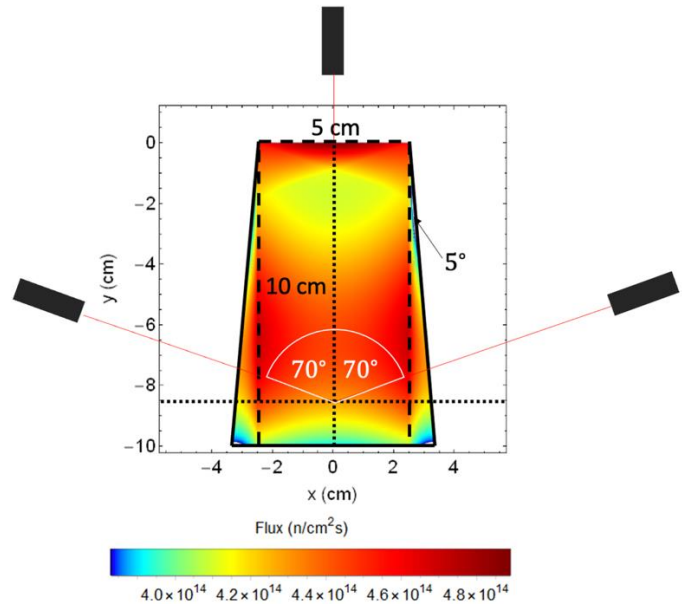

*Figure 14: Multi-beam flux flattening of the target station damage.*

**3. SUGGESTED TASKS ENABLING COMPACT CYCLOTRONS AS FPNS DRIVERS**

Figures 1, 2 and 3 provide insight into how the scale of compact cyclotrons may beneficially impact the economics of a US Fusion Prototypical Neutron Source. A first-order comparison on how a Cu-based or SC-based multi-driver system to the recently costed [28, 30] US IFMIF-lite scenario can be made. Assuming the development of 5-10mA individual Cu- or SC-based cyclotron drivers a summary of cost as compared to the IFMIF-lite is provided in the Table 2 below. This analysis takes advantage of the assumed higher duty cycle, the reduced need for installed power, and the capital cost of integrated beam current -on-target enabled by cyclotron drivers. Table 2 used the same analysis as presented by Egle and Ferguson [28, 30] and assuming green-field construction projects.

| Driver | Units | Energy MeV | Duty Cycle % | Integrated Current mA-yr | Power Use MW | Cost Per Accelerator M$ | Capital Cost of Integrated Beam-on-Target (M$/mA-yr) | Total Facility Cost M$ |
|---|---|---|---|---|---|---|---|---|
| RFQ/Linac IFMIF-lite | 1 | 40 | 65[+] | 81-88 | ~50 | 220-310[++] | 2.5-3.8 | 393-589[++] |
| Cu Cyclotron C-FPNS | 12 in facility | 35-40 | 92 facility | 50-100 | 5 (for 10) | < 11, turnkey | 1.32-2.64 | < 300[+] |
| SC Cyclotron C-FPNS | 12 in facility (10 in-use) | 35-40 | 92 facility | 50-100 | 2 (for 10) | < 10, turnkey | 1.2-2.4 | < 250[+] |

+Based on Knaster [1]  ++Based on Egle and Ferguson [28, 30]

*Table 2: Comparing the IFMIF-Lite concept, costed in [28, 30] with multiple Cu- and SC-based drivers.*

As expected (see figure 2), the footprint (and size) of superconducting (SC) cyclotrons are substantially less than their conventional (Cu) conducing competitors, and while the cost of current has been trending downward for all machines, a substantial and enabling cost reduction has occurred for SC cyclotrons over the past decade. A first-of-a-kind FPNS copper cyclotron driver provides current at approximately 5-10 mA at a cost of 1.5-3 M$/mA. This cost drops to less than 1-2 M$/mA for nth-of-a-kind (NOAK) machines. In comparison, the accelerator facility cost for a single IFMIF-like RFQ-Linac is approximately [28, 30] 220-393 M$ yielding a cost for current 1.8-3.4 M$/mA. While at face value the two systems provide beam current at approximately similar installed cost, the utility of applying multiple and relatively compact cyclotrons makes those systems more economically attractive. Specifically, the advertised availability of the IFMIF RFQ-Linac system is ~65%, while the commercial availability of cyclotrons for medical isotope production is >92%. Under an operating scenario where ten systems are kept on-line at all times and rotating-in machines through a schedule of maintenance, the overall facility availability would be dictated by non-driver components (i.e. target maintenance and sample removal, etc.) For Table 2, a cyclotron facility availability of 92% was assumed. Given these factors the cyclotrons then have a clear advantage in cost of integrated beam-on-target (cf. Table 2). Moreover, saving can be assumed both in terms of power and overall facility/building costs when utilizing cyclotrons. As approximately 40% of the facility cost is associated with the facility square footage in green-field, the compact cyclotron provides allow for a less costly build. Importantly, assuming such a significantly reduced facility foot print and power need can be realized, the potential for a less-costly *brown-field* construction is also possible.

| Milestone | Year 1 | | Year 2 | | Year 3 | | Year 4 | |
|---|---|---|---|---|---|---|---|---|
| | H1 | H2 | H1 | H2 | H1 | H2 | H1 | H2 |
| *Milestone 1:* *Milestone 1a:* Procurement of High-Current Commercial Ion Source and Development of Injection System Consistent with 35 MeV Cyclotron. Demonstration of injection of 5 mA of $H^{2+}$ and establishment of stable vortex at turn 6. | 0.75 | | 0.25 | | | | | |
| *Milestone 1b:* Demonstration of injection of 5 mA of $D^+$ and establishment of stable vortex at turn 6 | | | | | | | | |
| *Milestone 2a:* Detailed cyclotron physics and Engineering Drawings of First-of-a-kind high-current Cu-Cyclotron | 0.5 | | 0.8 | | | | | |
| *Milestone 2b:* Detailed cyclotron physics and Engineering Drawings of First-of-a-kind high-current SC-Cyclotron | 0.4 | | 0.4 | | 0.2 | | | |
| *Milestone 3:* *Provide materials and cyclotron physics optimizations consistent with a compact and maintainable cyclotron.* | 0.35 | | 0.35 | | | | | |
| *Milestone 4:* *Technoeconomic analysis or a C-FPNS system including potential siting and operational scenarios* | 0.3 | | 0.3 | | | | | |
| *Milestone 5:* *Construction and Operation of the FOAK C-FPNS Cyclotron.* | | | | | 5 | | 10 | |
| **Budget (M$)** | **2.25** | | **2.1** | | **5.2** | | **10** | |
| **Total (M$)** | **4.35** | | | | **15.2** | | | |

*Table 3: Schedule and budget of milestones leading to C-FPNS procurement.*

As discussed in Section 2, while the essential elements of high-current deuterium cyclotrons are in place and their commercial base exists, several key enabling technologies are at low technology readiness and their realization is uncertain. Here we suggest a focused research and development program to close the technical gap to near-term construction of a 5-10 mA, 35 MeV deuterium cyclotron by a leading commercial vendor. The technical milestones leading to a final procurement are provided in Table 3 along with an assumed schedule and budget. Detail tasks outlining these milestones follow.

*Task 1: High current beam injection:*

For both high-energy density physics experiments and medical isotope production the trend in cyclotron development has been towards axial injection, very high reliability, and currents exceeding 1 mA. Within this task we build upon current efforts driven by HEP particle physics needs and geared towards $H_2^+$ cyclotrons. As example, the current IsoDAR sterile neutrino project is on track to product a 5 mA $H_2^+$ compact cyclotron injector at end of calendar 2023. As proposed here a high-current, commercial $D^+$ source will be procured to replace the IsoDAR project $H_2^+$ source with a newly designed and constructed RFQ (cf. Section 2.2). A key beam physics issue to resolve will be the pre-acceleration and bunching of the deuterium beam using a custom-built radiofrequency quadrupole. As commercially available ion sources are available to very high current, the ultimate current injection limit will be set by the team's ability to design the completed injection system and test injection on an appropriate, but lower energy cyclotron.

> ***Milestone 1a: Procurement of High-Current Commercial Ion Source and Development of Injection System Consistent with 35 MeV Cyclotron. Demonstration of injection of 5 mA of $H^{2+}$ and establishment of stable vortex at turn 6.***
>
> ***Milestone 1b: Demonstration of injection of 5 mA of $D^+$ and establishment of stable vortex at turn 6.***

*Task 2: Critical physics of normal and superconducting cyclotrons:*

Even though 70 MeV proton/35 MeV deuteron cyclotrons are in commercial production, once the injector of Task 1 is introduced a range of cyclotron engineering and physics will be required to modify those *normal* conducting machines for the higher current including the collimation and extraction issues discussed in Section 2. Moreover, as no direct analog currently exists for a commercial 70 MeV superconducting cyclotron a higher level of design will be required for the high-current deuterium SC cyclotrons. This task interfaces with the commercial cyclotron vendors at the engineering and basic R&D level as well as for high-level simulation of the cyclotron physics performance.

Once axially injected into the cyclotron, the 5-10 mA $D^+$ beam will be refocused and brought into the median plane using a specially designed spiral inflector. Collimator central-region placement to be optimized for beam matching and clean-up using high fidelity modeling (cf. Section 2.2) Assumption of ~2 MeV fast acceleration using 4 double-gap RF cavities to increase inter-turn separation of orbits. Arcing near injection and space charge limitations are of primary concern in cyclotron physics simulations, though given the current high-current-density capability of proton cyclotrons and lesser degree of susceptibility of $D^+$ beams to space charge issues there is a high degree of confidence with regards to the normal conducting cyclotrons.

> ***Milestone 2a: In concert with Commercial Vendor, detail cyclotron physics of First-of-a-kind high-current Cu-Cyclotron and present in a ready-to-procure set of design drawings.***
>
> ***Milestone 2b: Milestone 2: In concert with Commercial Vendor, detail cyclotron physics of First-of-a-kind high-current SC-Cyclotron and present in a ready-to-procure set of design drawings.***

*Task 3: Relation of component activation to operation and safety:*

As discussed in Section 2, and given the use of high-current and high-energy deuterium beams, a level of activation of the cyclotron drivers, especially from collimators and septum materials. The level of allowed activation will be defined in collaboration with the detailed physics analysis of Task 2 and brought to an acceptable level through detailed physics and selection of alternative collimator/septum materials. The goal of this task is to define and realize an acceptable level of cyclotron activation with maintenance and facility shielding reduction being primary considerations.

> ***Milestone 3: Provide materials and cyclotron physics optimizations consistent with a compact and maintainable cyclotron.***

*Task 4: C-FPNS Siting, Cost Drivers and Economics*

Aspects of beam-target design including the use of multiple beams will be analyzed (cf. Section 2.5) to determine the optimal beam configuration yielding the desired metrics of annual fluence and maximum fluence gradient as proscribed in [27]. Potential beam-target configurations to be limited to those geometries consistent with future beam upgrades. With this information the required total facility beam current is determined and applied in the economic assessment. Further, it is noted that the Table 2 cost analysis was based on green-field construction at a national laboratory (ORNL). While it is either most beneficial, or required, to co-locate a FPNS facility with a hot-cell facility, a C-FPNS based on either Cu-based or SC-based cyclotrons do not necessarily require green-field construction. The potential for a lower-cost brown-field construction (cf. Section 2.5 and elsewhere) may be gained from a combined reduction in accelerator building foot-print, associated shops, and reduced power-required of cyclotrons as compared to a classic RFQ/Linac. In this task the original IFMIF-lite analysis will be expanded for both green-field and brown-field construction of the facility based on the engineering drawing realized through Task 2.

> ***Milestone 4: Technoeconomic analysis of a C-FPNS system including potential siting and operational scenarios.***

*Task 5: Construction and Initial Operation of First-of-a-Kind FPNS Cyclotron Driver*

As result of Task 2, engineering drawings of a C-FPNS driver (based on Cu or SC-cyclotrons) will be realized. Those drawings will be based on high-current D-beam technologies developed under the program and therefore not proprietary to any single cyclotron provider. With these engineering drawing and specifications, a Cu or SC-based cyclotron will be constructed by a commercial vendor. Beyond defining the cost and initial performance of the FOAK cyclotron, this machine will be installed for performance testing to understand critical performance parameters including current availability over time it will also be available to test and confirm critical beam-on-target parameters necessary for final facility design.

> ***Milestone 5: Construction and Operation of the FOAK C-FPNS Cyclotron.***

**4. TEAM, POTENTIAL OTHER COLLABORATORS, AND TIMELINESS OF RESOURCES**

***This is an opportune time to perform this research, as our proposal team can effectively leverage technical capabilities and ongoing collaborations with cyclotron (IsoDAR) and fusion materials (FUSMAT) communities.***

The current framework for the IsoDAR project provides a good template on how the cyclotron and irradiation facility tasks can be retired in anticipation of a near-term C-FPNS. In that project the academic team has taken leadership in defining performance and retiring technical risk, while engaging directly and strongly with the cyclotron industry on engineering physics aspects of machine design. Not only does this path highly leverage both the IsoDAR and fusion materials and technologies programs, it reduces the technical risk and burden placed on the cyclotron -industry-partner as they will be engaged in areas within their current technical portfolio. In other words, as these companies are quite motivated by the prospect of an advanced machine and would be very interested in eventually building a large number of them, they have a current focus on their core business and would be hesitant to directly lead such an R&D project.

It is currently envisioned that Tasks 1-3 (ion injection, collimation, and extraction) would be led by the MIT-based team currently involved in the IsoDAR project, with significant aspects of cyclotron engineering and physics (final cyclotron beam physics design, engineering design and drawing) carried out in close connection with a major commercial partner such as IBA or Best. Elements of Task 3 related to machine activation, materials selection, and target design will be carried out by a fusion technologies program group, including efforts from University participants such as Stony Brook and the University of Tennessee in consultation with the broader fusion community and the Oak Ridge National Laboratory with respect to the beam-on-target parameters testing. All tasks and performers would inform the technoeconomic analysis.